\documentclass[11pt,a4paper]{article}
\usepackage{amsfonts,amscd,amsmath,amssymb,amsthm}

\newtheorem{theo}{Theorem}[]

\newtheorem{ex}{Example}[]

\DeclareMathOperator{\Ker}{Ker}

\newcommand{\R}{\mathbb{R}}
\newcommand{\C}{\mathbb{C}}

\newcommand{\1}{\mathbb{I}}

\newcommand{\cD}{{\cal D}}
\newcommand{\cH}{{\cal H}}
\newcommand{\cM}{{\cal M}}

\newcommand{\cW}{{\cal W}}

\title{Kirchhoff's Rule for Quantum Wires. II: The Inverse Problem  with Possible Applications to Quantum Computers}
\author{V. Kostrykin\thanks{e-mail: kostrykin@t-online.de, kostrykin@ilt.fhg.de}\\
Lehrstuhl f\"ur Lasertechnik\\ Rheinisch
- Westf\"alische Technische Hochschule Aachen\\
Steinbachstra{\ss}e 15, D-52074 Aachen, Germany
\and
and\\
 R. Schrader\thanks{e-mail: schrader@physik.fu-berlin.de, Supported in part by
DFG SFB 288 ``Differentialgeometrie und Quantenphysik''}\\
Institut f\"{u}r Theoretische Physik\\
Freie Universit\"{a}t Berlin, Arnimallee 14\\
D-14195 Berlin, Germany}

\begin{document}

\maketitle

\begin{abstract}
In this article we continue our investigations of one particle quantum
scattering theory for Schr\"{o}dinger operators on a set of connected (idealized
one-dimensional) wires forming a graph with an arbitrary number of open ends.
The Hamiltonian is given as minus the Laplace operator with suitable linear
boundary conditions at the vertices (the local Kirchhoff law). In ``Kirchhoff's
rule for quantum wires'' [\textit{J.\ Phys.\ A: Math.\ Gen.} \textbf{32}, 595
-- 630 (1999)] we provided an explicit algebraic expression for the resulting
(on-shell) S-matrix in terms of the boundary conditions and the lengths of the
internal lines and we also proved its unitarity. Here we address the inverse
problem in the simplest context with one vertex only but with an arbitrary
number of open ends. We provide an explicit formula for the boundary conditions
in terms of the S-matrix at a fixed, prescribed energy. We show that any
unitary $n\times n$ matrix may be realized as the S-matrix at a given energy by
choosing appropriate (unique) boundary conditions. This might possibly be used
for the design of elementary gates in quantum computing. As an illustration we
calculate the boundary conditions associated to the unitary operators of some
elementary gates for quantum computers and raise the issue whether in general
the unitary operators associated to quantum gates should rather be viewed as
scattering operators instead of time evolution operators for a given time
associated to a quantum mechanical Hamiltonian. We also suggest an approach by
which the S-matrix in our context may be obtained from ``scattering
experiments'', another aspect of the inverse problem. Finally we extend our
previous discussion, how our approach is related to von Neumann's theory of
selfadjoint extensions.
\end{abstract}

\newpage

\section{Introduction}

 In this article we continue our investigation of quantum mechanical
scattering theory on a set of connected wires idealized to have zero diameter
and with an arbitrary number of open ends \cite{KS}. The dynamics is given by
the one-dimensional Laplacian with arbitrary boundary conditions at the
vertices, which makes it a selfadjoint operator. We view this as an idealized
version of a thin conducting wire with electrons in the conductance band, i.e.\
we have set $\hbar=2m^{*}=1$ with $m^{*}$ being the effective mass and we have
neglected the spin. Also if the wires are sufficiently thin, transversal modes
may be neglected and only longitudinal modes, i.e.\ motion in the direction of
the graph, should play a role. Below we will comment briefly on the associated
mathematical problem. A concrete example we have in mind are nanotubes, which
are considered to be interesting candidates for the design of microscopic
electronic devices (see e.g.\ \cite{Dekker-Nature98,PSSG,PSG} for recent,
possibly relevant experiments). In fact, the possibility of connecting
nanotubes of different diameter and chirality has created considerable interest
recently (see e.g.\ \cite{IIA,SDD,CEL,CCC,MS,Srivastava:98}). Alternatively
such graphs could be realized in terms of grooves etched on suitable surfaces
or in terms of strings of atoms forming a (planar) graph deposited again on a
suitable surface \cite{EZ,ZLE,SE}, see also  \cite{MNR,MZR}.

For our graph model we provided an explicit algebraic expression for the
S-matrix at a given energy in terms of the boundary conditions and the lengths
of the internal wires. We proved unitarity so this provides an example for the
Landauer-B\"{u}ttiker formalism \cite{La,Bue}, see also e.g.\ \cite{Datta} for an
extensive discussion. We note that the transmission rate for the connection of
two nanotubes with different diameter has been calculated in \cite{TT}. The
influence of disorder on the conductance characteristics of nanotubes was
studied in \cite{Anantram:98}.

In this article we will also address the inverse problem, namely the
determination of the boundary conditions given the S-matrix at a fixed energy.
We will only cover the case of one vertex only, since the general case with
internal wires and several vertices is much more difficult. Note that another
kind of inverse problems was considered previously in \cite{Gerasimenko}.

We will use this discussion of the one vertex case to indicate possible
applications in the design of quantum computers (see e.g.\
\cite{Preskill,Shor,Steane} and the literature quoted there). Recall that in
classical network theory one uses the theory of unitary matrices to describe
(deterministic) input-output relations and one even speaks of scattering
matrices (see e.g.\ \cite{Re1,Re2}). For quantum computers we advocate the idea
that one should use the theory of quantum scattering and the associated notion
of a unitary S-matrix to formulate (probabilistic) quantum input-output
relations. This differs from standard discussions on this subject in which the
unitary matrix is considered to be the unitary time evolution operator for a
given, fixed time and with a Hamiltonian as infinitesimal generator, which
describes the dynamics. Also we consider the rule by which the connection of
gates is supposed to correspond to matrix multiplication of the associated
unitary matrices not to be quite convincing. In fact, connecting two gates
means that one has a coupled system for which the relevance of the Hamiltonians
of the two subsystems remains unclear. So from the point of view of information
transmission, we consider it at least natural to ask for the relevance of
scattering matrices in the context of quantum computation. In fact, it is our
understanding that most of the present experimental designs for quantum gates
describe scattering experiments. In the present context of quantum wires the
incoming signals would be plane waves at a fixed energy in each of the wires
and which will be scattered at the vertex into outgoing plane waves again in
each of the wires and of the same energy.

In this context it is worthwhile to mention the connection between the
time dependent quantum mechanical description and scattering theory.
The latter describes the long time behaviour of quantum
evolution. Thus the use of
scattering matrices instead of unitary time evolution operators can only be
adequate provided the ``tact frequency'' of the quantum computer is
not too high and this depends on its physical realization. However, a
discussion of this issue as well as the question of how to incorporate
this proposed realization of quantum gates into the general context of
quantum computation with its notion of entanglement lies outside the
scope of the present article.

 The discussion below will be based on our result that in the
situation of a single vertex quantum wire any unitary matrix may be
uniquely realized as the S-matrix at a given energy by choosing
appropriate boundary conditions. In spirit this is similar to the
discussion in e.g.\ \cite{Reck} by
which any unitary matrix in principle can be realized with suitable beam
splitters and phase shifters. As an illustration we will calculate these
boundary conditions for some simple gates. We leave out the question of how to
implement these boundary conditions at the vertices of concrete physical wires.
However, in the present context we will discuss the general experimental
difficulty associated with designing quantum gates with a prescribed unitary
matrix. It has to do with the well known fact that usually only the absolute
values of matrix elements of the S-matrix are observable (conventionally given
in terms of cross sections). We will suggest an ``experimental'' prescription,
how one may also determine the phases of the S-matrix elements by suitable
experiments. This method may also be applied to potential scattering theory for
Schr\"{o}dinger operators \cite{KS2}.

We conclude this article by elaborating in more detail than in \cite{KS} on the
relation of our discussion of selfadjoint extensions with the corresponding
theory of von Neumann.

\section{The inverse problem for a quantum wire with a single vertex}

In this section we first discuss the inverse problem for $n$ (infinite) wires
ending at a single vertex, i.e.\ we will determine the boundary conditions
given the S-matrix. We briefly recall the set-up for this situation. Let the
Hilbert space be given as
\begin{equation*}
\cH=\oplus^{n}_{i=1}\cH_{i}=\oplus^{n}_{i=1}L^{2}([0,\infty)).
\end{equation*}
Intuitively the $n$ origins $0$ are to be identified. This will be achieved
in a moment by describing the quantum mechanical dynamics.
Elements $\psi\in\cH$ will be written as
$\psi=(\psi_{1},\psi_{2},...,\psi_{n})$ and derivatives as
$\psi^{\prime}=(\psi_{1}^{\prime},\psi_{2}^{\prime},...,\psi_{n}^{\prime})$
 We will
call $\psi_{j}$ the component of $\psi$ in channel $j$. The scalar product in $\cH$ is
\begin{equation*}
 \langle\phi,\psi\rangle=\sum^{n}_{i=1}\langle\phi_{i},\psi_{i}\rangle_{\cH_i}
\end{equation*}
with the standard scalar product on $L^{2}([0,\infty))$ on the right hand side.
Now the dynamics will be given by the Laplace operator away from the origin
supplemented by suitable boundary conditions at the origin. For this we proceed as follows.
Consider the symmetric operator $\Delta^0$ on $\cH$, such that
\begin{displaymath}
\Delta^0\psi=\left(-\frac{d^2\psi_1}{dx^2},\ldots,-\frac{d^2\psi_n}{dx^2}\right)
\end{displaymath}
with domain of definition $\cD (\Delta^{0})$ being the set of all
$\psi$ with $\psi,\psi^{\prime},\psi^{\prime\prime}\in \cH$ and such
that
$\psi_{i}(0)=\psi_{i}^{\prime}(0)=0$ for all $1\le i\le n$.
All selfadjoint extensions are now given as follows. Let
$\cD\subset\cH$ be the set of all $\psi$ such that
$\psi,\psi^{\prime},\psi^{\prime\prime}$ are all in
$\cH$.
 On $\cD$ consider the following skew-Hermitian
quadratic form given as
\begin{equation*}
 \Omega(\phi, \psi)= \langle\Delta\phi,\psi\rangle-\langle\phi,\Delta\psi\rangle
=-\overline{\Omega(\psi,\phi)}
\end{equation*}
with the Laplace operator $\Delta=-d^2/dx^2$ considered as a differential
operator. Obviously $\Omega$ vanishes identically on $\cD(\Delta^{0})$. Any
self-adjoint extension of $\Delta^{0}$ is now given in terms of a maximal
isotropic subspace of $\cD$, i.e.\ a maximal (linear) subspace on which
$\Omega$ vanishes identically. To find these maximal isotropic subspaces, let $
[\:]: \cD\rightarrow \C^{2n}$ be the surjective linear map which associates to
$\psi$ and $\psi^{\prime}$ their boundary values at the origin:
\begin{equation*}
[\psi]=(\psi_{1}(0),...\psi_{n}(0),
       \psi^{\prime}_{1}(0),...\psi^{\prime}_{n}(0))^{T}=
   \left( \begin{array}{c}\psi(0)\\
                          \psi^{\prime}(0) \end{array} \right).
\end{equation*}
Here $T$ denotes the transpose, so $[\psi],\psi(0)$ and $\psi^{\prime}(0)$ are
 considered to be column vectors of length $2n$ and $n$ respectively.
The kernel of the map $[\:]$ is obviously equal to
$\cD(\Delta^{0})$. Then we have
\begin{equation*}
\Omega(\phi,\psi)=\omega([\phi],[\psi]):=\langle[\phi],J[\psi]\rangle_{\C^{2n}},
\end{equation*}
where $\langle\,,\,\rangle_{\C^{2n}}$ now denotes the scalar product on $\C^{2n}$ and
where the $2n\times 2n$ matrix $J$ is the canonical symplectic matrix on $\C^{2n}$:
\begin{equation}
\label{1}
 J= \left ( \begin{array}{cc}0&\1\\
                            -\1&0 \end{array} \right).
\end{equation}
Here and in what follows $\1$ is the unit matrix for the given context.

To find all maximal isotropic subspaces in $\cD$ with respect to $\Omega$ it
therefore
suffices to find all maximal isotropic subspaces in $\C^{2n}$ with respect to $\omega$ and
to take their preimage under the map $[\:]$.
Let the linear subspace $\cM=
\cM (A,B)$ of
$\C^{2n}$ be given as the set of all ${[\psi]}$ in $\C^{2n}$ satisfying
\begin{equation}
\label{2}
 A\psi(0)+B\psi^{\prime}(0)=0,
\end{equation}
where $A$ and $B$ are two $n\times n$ matrices. If the $n\times 2n$ matrix
$(A,B)$ has maximal rank equal to $n$ and if $AB^{\dagger}$ is selfadjoint then
$\cM(A,B)$ is maximal isotropic and in this way one obtains all maximal
isotropic subspaces in $\C^{2n}$. Two maximal isotropic subspaces
$\cM(A^{\prime},B^{\prime})$ and $\cM(A,B)$ are equal iff there is an
invertible $n\times n$ matrix $C$ such that $(A^{\prime},B^{\prime})=(CA,CB)$.
In this sense we will say that the $n\times 2n$ matrix $(A,B)$ is uniquely
fixed by the boundary condition. Let $\cD(A,B)$ denote the inverse image of
$\cM(A,B)$ under the map $[\:]$. Then
$\cD(\Delta^{0})\subset\cD(A,B)\subset\cD$ and $\Delta \upharpoonright\cD(A,B)$
is the core for a selfadjoint extension of $\Delta^{0}$ and which is denoted by
$\Delta(A,B)$. This one-dimensional set-up could be a limiting case of a more
realistic set-up with wires having non zero diameter. More precisely consider a
graph to be a fixed subset of $\R^{3}$ such that in particular the angles
between the edges ending at any vertex are fixed. Now form a tubular
$\epsilon$-neighborhood of this graph, i.e.\ a real system of connected wires
of diameter $\epsilon$. In mathematical terms this means that the graph is a
deformation retract of any of these tubular neighborhoods. Consider the Laplace
operator on this 3-dimensional set with Neumann boundary conditions its
boundary, i.e.\ on its surface. Now the mathematical question arises whether
for $\epsilon\downarrow 0$ the resolvent (i.e. the Green's function) converges
suitably to the resolvent of a Laplace operator on the graph and which is of
the above type. If this is the case its boundary conditions must then be given
in terms of geometrical data, i.e. on how this graph lies in $\R^{3}$. In other
words and as already mentioned in the introduction, when $\epsilon\downarrow 0$
the transversal modes (except $0$) should tend to infinity while the
``longitudinal'' modes should converge to those of a suitable Laplace operator
on the graph. For some simple cases this is indeed known to be true (see for
example \cite{MSch}). The S-matrix $S(E)=S_{A,B}(E)$ for any fixed energy $E>0$
is now given as follows. It is an $n\times n$ matrix whose elements are defined
by the following relations. Consider the plane wave solutions
$\psi^{k}(\cdot,E)$, $1\le k\le n$ of the form
\begin{equation}
\label{3}
 \psi^{k}_{j}(x,E)=\delta_{jk}e^{-i\sqrt{E}x}+S_{jk}(E)e^{i\sqrt{E}x}
\end{equation}
and which satisfy the boundary conditions. Then the S-matrix at energy
$E$ for the
boundary condition $(A,B)$ is given as
\begin{equation}
\label{4}
 \begin{array}{ccc}S_{A,B}(E)&=& -(A+i\sqrt{E}B)^{-1}(A-i\sqrt{E}B)\\
   {}&   =&-(A^{\dagger}-i\sqrt{E}B^{\dagger})(AA^{\dagger}+EBB^{\dagger})^{-1}
  (A-i\sqrt{E}B)\end{array}
\end{equation}
and is unitary and real analytic in $E>0$. Note that
$S_{A,B}(E)^{-1}=S_{A,-B}(E)$. In \cite{KS} it was shown that the knowledge of
the S-matrix at a fixed energy, $E_{0}$ say, uniquely fixes the boundary
conditions. Here we will provide an explicit  solution to this inverse problem.
For known $S=S_{A,B}(E_{0})$ let
\begin{equation}
\label{5}
 A^{\prime}=-\frac{1}{2}(S-\1),\qquad B^{\prime}=\frac{1}{2i\sqrt{E_{0}}}(S+\1).
\end{equation}
Then by \eqref{4} one has
$(A^{\prime},B^{\prime})=(CA,CB)$ with $C=(A+i\sqrt{E_{0}}B)^{-1}$.
This proves that $S=S_{A,B}(E_{0})$ indeed uniquely fixes the boundary
condition. We can even say more. Indeed, observe that $(A^{\prime},B^{\prime})$
defined by \eqref{5}  satisfies the two relations
\begin{equation*}
A^{\prime}B^{\prime\dagger}=\frac{1}{2i\sqrt{E_{0}}}(S^{\dagger}-S),\qquad
A^{\prime}+i\sqrt{E_{0}}B^{\prime}=\1,
\end{equation*}
where the first relation is a simple consequence of unitarity. From the first
relation we can also directly read off that $A^{\prime}B^{\prime\dagger}$ is selfadjoint
and from the second that $(A^{\prime},B^{\prime})$
has maximal rank equal to $n$. Thus we have proved the following
\begin{theo}
For any unitary $n\times n$ matrix $S$ and any energy $E_{0}>0$
there are unique boundary conditions $(A,B)$ such that
$S_{A,B}(E_{0})=S$. More generally the relation
\begin{equation}
\label{7}
\begin{array}{ccc}
S_{A,B}(E)&=&\left((\sqrt{E}-\sqrt{E_{0}})S
              +(\sqrt{E}+\sqrt{E_{0}})\right)^{-1}\\
         &&\cdot\left((\sqrt{E}+\sqrt{E_{0}})S+
         (\sqrt{E}-\sqrt{E_{0}})\right)
\end{array}
\end{equation}
holds for all $E>0$.
\end{theo}
Indeed, the relation \eqref{7} follows from
$S_{A,B}(E)=S_{A^{\prime},B^{\prime}}(E)$ for all $E>0$ and from relation
\eqref{4}.
Since $E_{0}$ was arbitrary this in particular gives the relation
\begin{equation}
\label{8}
\begin{array}{ccc}
S_{A,B}(E)&=&\left((\sqrt{E}-\sqrt{E_{0}})S_{A,B}(E_{0})
              +(\sqrt{E}+\sqrt{E_{0}})\right)^{-1}\\
         &&\cdot\left((\sqrt{E}+\sqrt{E_{0}})S_{A,B}(E_{0})
+(\sqrt{E}-\sqrt{E_{0}})\right)
\end{array}
\end{equation}
for all $E$ and $E_{0}>0$. In particular this relation may be used as
a test to verify if the S-matrix indeed results from boundary
conditions in the way described here.

As an illustration with possible applications to quantum computers in mind we
consider the some elementary gates usually discussed in this context (see e.g.\
\cite{Bar}). Thus we will view a vertex with $n$ wires entering as a quantum
gate with the wires viewed as channels. In particular the output channels are
the same as the input channels. Therefore in
addition to quantum transmission from one channel into a different one there
will also be reflection from a channel into itself. These reflection amplitudes
correspond to the notion of backpropagation in classical information theory.

It is important to realize that here we are dealing with a one-particle theory.
In fact, the notion of a particle is introduced here for the following reason.
In the concrete setup of classical computers one deals with signals, which are
localized in space and time. But in quantum (field) theory and with the
appropriate modifications this is one of the notions used to describe particles
in terms of wave packets. In the quantum mechanical formulation incoming
signals at gates result in outgoing signals and this corresponds to looking at
scattering theory and the associated unitary S-matrix. In particular in our
approach the vertices act like classical external potentials for which the
associated quantum scattering theory has been discussed extensively (see e.g.\
\cite{RS1}). Here the dimension of the space of incoming plane waves at energy
$E$ is equal to $n$, the number of wires entering the vertex. This contrasts
with the usual discussion of quantum gates with $n$ input and $n$ (different)
output channels, where the Hilbert space has dimension equal to $2^{n}$ and
replaces the set of classical information containing $n$ bits. This dimension
reflects the tensor product structure ($n$ factors, each being two-dimensional)
of the underlying theory. The tensor product structure is essential for the
notion of entanglement, the main ingredient for quantum computation. It is
entanglement makes the drastic difference as compared with classical
computation. The exponential dependence of the dimension on $n$ is another
reason for the attractiveness of quantum computing. The underlying picture in
this case is therefore given by a multi-particle quantum theory, recall for
example the discussion of the Einstein-Podolsky-Rosen paradox. A reconciliation
of our discussion with the tensor product structure will need further
investigations but again, our main motivation was to propose a possible
physical realization of unitary operators associated to elementary quantum
gates.

Our approach may also be adapted to the situation where we consider
nonrelativistic electrons with spin moving through the wires and with boundary
conditions possibly allowing for spin flips. Now for $n$ wires the Hilbert
space of incoming plane waves at fixed energy will have dimension equal to
$2n$. As mentioned in the introduction we will not discuss the experimental
feasibility of designing such vertices with a given boundary condition. Also we
will not discuss to what extent the dynamics given by the Laplace operator
$\Delta(A,B)$ may serve as an approximation to the dynamics of a real particle
(like an electron, say) in concrete wires, where for example one is also
confronted with the problem of localization and the resulting exponential decay
in space of some eigenfunctions due to the presence of impurities (see e.g.\
\cite{Landauer} for a discussion on this point) and which may prevent the
transmission (see \cite{Kostrykin:Schrader:preprint}).

\begin{ex}
In what follows the energy $E$ will be fixed at $E_{0}$. Consider the
one-dimensional unitary matrix given as a phase $\exp (i\chi)\; (n=1)$. Then
the Robin boundary condition $\cos\phi\;\psi(0)+\sin\phi\;\psi^{\prime}(0)=0$
with $\phi$ chosen such that
\begin{equation*}
e^{i\chi}=-\frac{\cos\phi -i\sqrt{E_{0}}\sin\phi}
               {\cos\phi +i\sqrt{E_{0}}\sin\phi}
\end{equation*}
solves this problem for one wire.
Next consider the $2\times 2$ unitary, idempotent so-called Hadamard matrix
$(n=2)$
\begin{equation*}
      \frac{1}{\sqrt{2}}\left( \begin{array}{cc}1&1\\
                        1&-1\end{array}\right).
\end{equation*}
Then the boundary conditions are given by
\begin{eqnarray*}
A&=&\frac{1}{2}
\left(\begin{array}{cc} 1-\frac{1}{\sqrt{2}}&-\frac{1}{\sqrt{2}}\\
               -\frac{1}{\sqrt{2}}&1+\frac{1}{\sqrt{2}}\end{array}\right),\\
B&=&\frac{1}{2i\sqrt{E_{0}}}
\left(\begin{array}{cc}1+\frac{1}{\sqrt{2}}&\frac{1}{\sqrt{2}}\\
                     \frac{1}{\sqrt{2}}&1-\frac{1}{\sqrt{2}}\end{array}\right).
\end{eqnarray*}
Note that both $A$ and $B$ have vanishing determinant.

Finally consider the real, unitary $4\times 4$ matrix (the XOR gate, which is a
special Toffoli gate, also called a controlled-NOT (``CNOT'')) $(n=4)$
\begin{equation*}
\left(\begin{array}{cccc} 1&0&0&0\\
                          0&1&0&0\\
                          0&0&0&1\\
                          0&0&1&0\end{array}\right).
\end{equation*}
The corresponding boundary conditions are now given as
\begin{eqnarray*}
A&=&\left(\begin{array}{cccc} 0&0&0&0\\
                              0&0&0&0\\
                              0&0&\frac{1}{2}&-\frac{1}{2}\\
                         0&0&-\frac{1}{2}&\frac{1}{2}\end{array}\right),\\
B&=&\frac{1}{2i\sqrt{E_{0}}}\left(\begin{array}{cccc}2&0&0&0\\
                                                     0&2&0&0\\
                                                     0&0&1&1\\
                                         0&0&1&1\end{array}\right).
\end{eqnarray*}
Again both $A$ and $B$ have vanishing determinant.
\end{ex}

The S-matrix at a given energy and resulting from connecting such (elementary)
gates is obtained from the S-matrices at the same energy of these individual
gates and the lengths connecting them by what we called the generalized star
product \cite{KS,KSK}, because it generalizes the star product in classical
network theory (see \cite{Re1,Re2}). Although associative this composition rule
is nonlinear due to the presence of reflection amplitudes. In the context of
potential (quantum) scattering theory on the line an equivalent formulation of
the star product also figures under the name Aktosun formula (see e.g.\ \cite{
Aktosun,Aktosun:Klaus:Mee,Bianchi,Bianchi2,Kostrykin:Schrader:99a}) and there
it is an easy consequence of the multiplicative property of the fundamental
solution of the Schr\"{o}dinger equation for a given energy and which is a $2\times
2$ matrix (see \cite{Kostrykin:Schrader:99a}). In the theory of mesoscopic
systems and multichannel conductors this has also been known for a long time
and there it is called the multiplicativity of the transfer matrix, which is
conjugate to the fundamental solution (see e.g.\ \cite{Tong,Dorokhov:82,
Dorokhov:83,Dorokhov:84,Dorokhov:88,Mello:Pereyra:Kumar,
Stone:Mello:Muttalib:Pichard,Beenakker,Datta})). In this context we would also
like to mention that the problem of how to connect gates in a concrete way
without feedback problems (back propagation) due to the presence of reflection
amplitudes seems to be well known (see e.g.\ \cite[p.\ 3458]{Bar},
\cite{Landauer}) but to the best of our knowledge as yet no concrete proposal
has been made, even on the conceptual level. Thus so far ordinary matrix
multiplication has been used as the composition rule when connecting several
gates (see e.g.\ \cite{Bar} and the references quoted there). This is in
analogy to the notion of straightline programming in information theory.

\section{Experimental determination of the S-matrix}

In this section we will address the general problem to what extent scattering
experiments fix the S-matrix. In what follows the quantum wire will be
arbitrary, i.e.\ it may contain an arbitrary number of external and internal
lines and with an arbitrary number of vertices (see \cite{KS} for details).
 First we recall that the matrix elements $S_{jk}(E)$ introduced in
\eqref{3} have the physical interpretation of transmission coefficients
$(j\neq k)$ and reflection coefficients $(j=k)$ into channel $j$ for an
incoming plane wave of energy $E$ in channel $k$. Hence their absolute values
are observable. On the other hand if one prepares an incoming wave packet
$\phi(\cdot;E)$ to be of the form
\begin{equation*}
     \phi_{k}(x;E)=\lambda_{k}e^{-i\sqrt{E}x}
\end{equation*}
with arbitrary complex amplitudes $\lambda_{k}\in \C$ ($1\leq k\leq n$), then
the resulting transmission amplitudes into channel $j$ are of the form
\begin{equation}
\label{11}
    \sum_{k=1}^{n}\lambda_{k}S_{jk}(E),
\end{equation}
whose absolute values in principle are observable. But an easy argument implies
the following. By a relabeling of the channels if necessary the
S-matrix may be decomposed into a direct sum as
\begin{equation}
\label{decom}
S_{A,B}(E)=\oplus_{m=1}^{M}S^{m}_{A,B}(E)
\end{equation}
for all $E>0$, where $M$ is supposed to be maximal with $1\leq M\leq n$. Here
the $S^{m}_{A,B}(E)$ are unitary $k_{m}\times k_{m}$ matrices with $\sum
k_{m}=n$. Indeed, since $S_{A,B}(E)$ is real analytic in $E>0$ any of its
matrix elements is either identically equal to zero or vanishes on at most a
denumerable set without accumulation points. This implies that such a maximal
decomposition \eqref{decom} exists and is unique. From the point of view of
boundary conditions this means that $(A,B)$ may be chosen to have a
corresponding maximal decomposition. This means that actually one has a
disconnected graph (see \cite{KS} for further details). Now experiments fix
$S_{A,B}(E)$ up to a phase factor matrix $\exp\,(i\underline{\chi}(E))$, where
\begin{equation}
\label{phase}
\exp\,(i\underline{\chi})(E)=\oplus_{m=1}^{M}\exp\,(i\chi_{m}(E))\1.
\end{equation}
So assume now that experimentally one has measured the S-matrix for all
energies in the form $\widetilde{S}(E)$ which agrees with $S_{A,B}(E)$ up to a
phase factor matrix $\exp\,(i\underline{\chi}(E))$, i.e.\ the relation
$S_{A,B}(E)=\exp\,(i\underline{\chi}(E))\widetilde{S}(E)$ holds for as yet
undetermined phase factors $\exp\,(i\chi_{m}(E)),\,1\leq m\leq M$. In
particular note that the decomposition \eqref{decom} is observable, i.e.\
$\widetilde{S}(E)$ has a decomposition of the same form. Since we know that
$S_{A,B}(E)$ is real analytic in $\sqrt{E}>0$ we may as well assume
$\widetilde{S}(E)$ to be real analytic by fitting with a phase factor matrix if
necessary. But then the phase factor matrix $\exp\,(i\underline{\chi}(E))$ is
also real analytic. In particular if the graph has one vertex only then by
\eqref{8} this results in the following relation for the phase factor matrix
\begin{eqnarray}\label{12}
\exp\,(i\underline{\chi}(E))&=&\widetilde{S}(E)^{-1}\left(\sqrt{E}-\sqrt{E_{0}})
              \exp\,(i\underline{\chi}(E_{0}))\widetilde{S}(E_{0})
              +(\sqrt{E}+\sqrt{E_{0}})\right)^{-1}\nonumber\\
 &&\cdot\left((\sqrt{E}+\sqrt{E_{0}})\exp\,(i\underline{\chi}(E_{0}))\widetilde{S}(E_{0})
        +(\sqrt{E}-\sqrt{E_{0}})\right)
\end{eqnarray}
for all $E$ and $E_{0}$. Now fix $E_{0}$. If there is a solution, then
 \eqref{12} shows
that $\exp\, (i\underline{\chi}(E))$ is completely determined in terms of the
``initial condition'' $\exp\,(i\underline{\chi}(E_{0}))$ for all $E$. On the
other hand in order to find a solution one has to choose
$\exp\,(i\underline{\chi}(E_{0}))$ in such a way that the r.h.s.\ of \eqref{12}
is a phase factor matrix for all $E>0$ in the sense of
\eqref{phase}. However, this procedure does not lead to a unique solution
in general as may be seen from looking at Robin boundary conditions
given as
\begin{equation*}
A_{jk}=\delta_{jk}\cos\phi_{k},\:B_{jk}=\delta_{jk}\sin\phi_{k}
\end{equation*}
resulting in an S-matrix of the form
\begin{equation*}
S_{jk}(E)=-\delta_{jk}\frac{\cos\phi_{k}-i\sqrt{E}\sin\phi_{k}}
           {\cos\phi_{k}+i\sqrt{E}\sin\phi_{k}}.
\end{equation*}
Then one can essentially only achieve
\begin{equation*}
\widetilde{S}_{jk}(E)=\delta_{jk}
\end{equation*}
by experiments and any initial condition leads to a solution. This general lack
of uniqueness is reminiscent of a similar situation in scattering theory
(\cite{Chrichton,Newton,Martin1,IzM,Martin2}, see also e.g.\ \cite {Martin}).

However, there is a way out to find the phases of the S-matrix. The idea is to
enlarge the graph by components for which the associated S-matrices
$S^{\mathrm{aux}}(E)$ are supposed to be known. The enlarged graph will give
rise to an S-matrix $S^{\mathrm{new}}(E)$, which is obtained from the original
S-matrix $S(E)$, $S^{\mathrm{aux}}(E)$ and the lengths of the new internal
lines (which also are supposed to be known) via the generalized star product
\cite{KS,KSK}. By varying $S^{\mathrm{aux}}(E)$ and measuring the absolute
values of the matrix elements of $S^{\mathrm{new}}(E)$, we may infer the phases
of the matrix elements of $S(E)$. This strategy is inspired by a familiar
procedure in quantum computation, where one analyzes the effect of a black box
(an ``oracle'') by combining the black box with another box, whose effect is
known (see e.g.\ \cite{Preskill} and the references quoted there).

We start with the case where we want to determine the phase of the reflection
amplitude $S_{ii}(E)$. If $S_{ii}(E)=0$, there is nothing to prove, so we will
assume $S_{ii}(E)\neq 0$. On the external line labeled by $i$ we introduce an
additional vertex at a distance $a_{i}$ from the vertex where the line ends.
There we introduce arbitrary boundary conditions which result in an arbitrary
$2\times 2$ unitary S-matrix $S^{\mathrm{aux}}(E)$ by our previous discussion.
By the techniques of the generalized  star product the resulting S-matrix
$S^{\mathrm{new}}(E)$ is obtained from $S(E)$, $S^{\mathrm{aux}}(E)$ and the
length $a_{i}$ such that in particular
\begin{equation}
\label{Snew}
S^{\mathrm{new}}_{ii}(E)=U_{22}+U_{21}S_{ii}(E)(1-S_{ii}(E)U_{11})^{-1}U_{12}.
\end{equation}
Here the $U_{kl},\;k,l=1,2$ are the matrix elements of the unitary
$2\times 2$ matrix
\begin{equation*}
U=\left(\begin{array}{cc}e^{i\sqrt{E}a_{i}}&0\\
                         0&1\end{array}\right)S^{\mathrm{aux}}(E)
  \left(\begin{array}{cc}e^{i\sqrt{E}a_{i}}&0\\
                         0&1\end{array}\right).
\end{equation*}
In the article \cite{KS} using a ``Born series''  expansion we explain the
physical intuition behind the generalized star product, from which in
particular \eqref{Snew} may easily be obtained. In \cite{KSK} we shall provide
a rigorous proof that the generalized star product indeed provides the right
factorization. Since $S^{\mathrm{aux}}(E)$ and $a_{i}$ are supposed to be
known, $U$ is also known and may be chosen arbitrary in the unitary group
$\mathsf{U}(2)$. Thus $S^{\mathrm{new}}_{ii}(E)$ is a function of $S_{ii}(E)$
and $U$ and by measurements $|S^{\mathrm{new}}_{ii}(E)|$ and $|S_{ii}(E)|$ are
known. Actually $S_{ii}^{\mathrm{new}}(E)=\exp (2i\sqrt{E}a_{i})\;S_{ii}(E)$
for the special case when
\begin{equation*}
S^{\mathrm{aux}}(E)=S_{0}=\left(\begin{array}{cc}0&1\\
                                  1&0\end{array}\right),
\end{equation*}
the ``unit matrix'' with respect to the generalized star product.

Write $S_{ii}(E)=\exp (i\phi)\;|S_{ii}(E)|$. Thus
\begin{equation*}
|S^{\mathrm{new}}_{ii}(E)|=\left|U_{22}+U_{21}|S_{ii}(E)|e^{i\phi}(1-|S_{ii}(E)|e^{i\phi}U_{11})^{-1}
               U_{12}\right|
\end{equation*}
is known for all $U$ and the aim is to determine $\phi$. Now any
$U\in\mathsf{U}(2)$ may be written as
\begin{equation*}
U= \left(\begin{array}{cc}e^{i(\chi+\tau)}\rho
      &-e^{i(\chi-\kappa)}(1-\rho^{2})^{1/2}\\
        e^{i(\chi+\kappa)}(1-\rho^{2})^{1/2}&e^{i(\chi-\tau)}\rho\end{array}
   \right)
\end{equation*}
with $0\le \rho\le 1$. Therefore
\begin{equation}
\label{max}
 \left|\rho- (1-\rho^{2})|S_{ii}(E)|
 (e^{-i(\phi+\chi+\tau)}-|S_{ii}(E)|\rho)^{-1}\right|
\end{equation}
is known. Now choose $\rho$ so small that
$2\rho^{2}|S_{ii}(E)|<|S_{ii}(E)|-\rho$. Then \eqref{max} is easily seen to be
maximal for $\exp (i(\phi+\chi+\tau))=-1$. Thus tuning $\chi$ and $\tau$ and
hence $S^{\mathrm{aux}}(E)$ while keeping $\rho$ fixed, such that
$|S^{\mathrm{new}}_{ii}(E)|$ becomes maximal, fixes $\phi$ as was claimed. An
alternative way to determine $\phi$ is to consider
\begin{equation*}
S^{\mathrm{new}}_{ij}(E)=S_{ij}(E)(1-S_{ii}(E)U_{11})^{-1}U_{12}
\end{equation*}
with $i\neq j$ under the same situation. This procedure therefore is only possible if the
graph considered has at least two external lines. Now
\begin{equation*}
|S^{\mathrm{new}}_{ij}(E)|=|S_{ij}(E)|(1-\rho^{2})^{1/2}
      \left|e^{-i(\phi+\chi+\tau)}-|S_{ii}(E)|\rho\right|^{-1}
\end{equation*}
becomes maximal when $\exp (i(\phi+\chi +\tau))=1$ and minimal when $\exp
(i(\phi+\chi+\tau))=-1$.

We turn to a discussion of the phases of the transmission amplitudes
$S_{ij}(E),\; i\neq j$. As discussed above (see \eqref{11}), by a suitable
preparation of the incoming state as a superposition of plane waves of energy
$E$ in the incoming channels $i$ and $j$, we can measure
$|S_{ij}(E)+\lambda\;S_{ii}(E)|$ for all complex $\lambda$. Since $S_{ii}(E)$
is has already been determined, this fixes $S_{ij}(E)$ for $i\neq j$ provided
$S_{ii}(E)\neq 0$. Now assume $S_{ii}(E)=0$. Then we proceed as follows. By the
same procedure as above we insert a vertex on the line labeled by $i$ at a
distance $a^{\prime}_{i}$ with an S-matrix $S^{\mathrm{aux}}(E)^{\prime}$. In
analogy to
\eqref{Snew} we then obtain an S-matrix $S(E)^{\prime}$ with
$S_{ii}(E)^{\prime}=U_{22}^{\prime}=S^{\mathrm{aux}}_{22}(E)^{\prime}$, which
is non vanishing provided $\rho^{\prime}>0$. Then we can measure
$S_{ij}(E)^{\prime}$ with the arguments used above. Now $S(E)_{ij}^{\prime}$
converges to $\exp (i\sqrt{E}a_{i}^{\prime})\;S_{ij}(E)$ when
$S^{\mathrm{aux}}(E)^{\prime}$ converges to $S_{0}$, so this determines
$S_{ij}(E)$. The appearance of the phase factor $\exp
(i\sqrt{E}a_{i}^{\prime})$ is due to the fact that the insertion of the new
vertex shifts the origin of the line labeled by $i$ by the distance
$a^{\prime}_{i}$. This concludes our proof that all matrix elements $S_{ij}(E)$
may be obtained from scattering experiments performed in this way. Note that
for this procedure only scattering experiments at fixed energy are necessary.

\section{Description in terms of von Neumann's extension theory}

With the observations of the Section 2 we also may describe
$\Delta(A,B)$ defined on a graph with a single vertex and the
resulting S-matrix from the viewpoint of von Neumann's extension
theory (see e.g.\ \cite{RS}). This extends the discussion in
Appendix A of \cite{KS}. According to von Neumann's theorem any
self-adjoint extension $\Delta$ of $\Delta^0$ can be uniquely
parameterized by a linear isometric isomorphism
$\cW:\Ker(-{\Delta^0}^\dagger-i)\rightarrow\Ker(-{\Delta^0}^\dagger+i)$
according to the formula
\begin{eqnarray*}
&&\cD(\Delta) = \left\{\psi+\psi_++\cW\psi_+|\ \psi\in\cD(\Delta^0),
\psi_+\in\Ker(-{\Delta^0}^\dagger-i) \right\},\\
&&-\Delta(\psi+\psi_++\cW\psi_+)=-\Delta^0\psi+i\psi_+-i\cW\psi_+.
\end{eqnarray*}
We will describe $\cW$ in terms of a matrix $W$ by choosing
particular bases in $\Ker(-{\Delta^0}^\dagger-i)$ and
$\Ker(-{\Delta^0}^\dagger+i)$ respectively.

Namely we choose $u_j\in\Ker(-{\Delta^0}^\dagger-i)$ and
$v_j\in\Ker(-{\Delta^0}^\dagger+i)$, $j=1,\ldots,n$ to be given as
\begin{equation*}
(u_j(x))_k=\delta_{jk}2^{1/4}e^{\frac{1}{\sqrt{2}}(-1+i)x},\quad
(v_j(x))_k=\delta_{jk}2^{1/4}e^{\frac{1}{\sqrt{2}}(-1-i)x}, \quad k=1,\ldots,n.
\end{equation*}
One can easily verify that $\{u\}_{j=1}^n$ and $\{v\}_{j=1}^n$ are orthonormal bases for
$\Ker(-{\Delta^0}^\dagger-i)$ and $\Ker(-{\Delta^0}^\dagger+i)$, respectively.
Thus
$W$ is the unitary matrix representation of $\cW$ with respect to
these bases
$\{u\}_{j=1}^n$ and $\{v\}_{j=1}^n$, i.e.
\begin{displaymath}
\cW u_j=\sum_{k=1}^{n}W_{kj}v_{k}.
\end{displaymath}
In Appendix A of \cite{KS} it was shown that for $\Delta=\Delta(A,B)$ the matrix
$W=W_{A,B}$ was given as
\begin{equation*}
W_{A,B}=-\left(A-\frac{1}{\sqrt{2}}(1+i)B\right)^{-1}
         \left(A+\frac{1}{\sqrt{2}}(-1+i)B\right).
\end{equation*}
But this means that given any unitary matrix $W$ describing a selfadjoint
extension in the sense of von Neumann in the basis described
above, and using the pair $(A^{\prime},B^{\prime})$
given as $(A^{\prime},B^{\prime})=(CA,CB)$ with
$C=\sqrt{2}(A-1/\sqrt{2}(1+i)B)^{-1}=\sqrt{2}(A-\exp(i\pi/4)B)^{-1}$
to describe the boundary conditions, we have $W=-\1-iB^{\prime}$ as
well as $W=\1-\sqrt{2}A^{\prime}+B^{\prime}$ giving $A^{\prime}$ and
$B^{\prime}$ in terms of $W$ as
\begin{eqnarray*}
A^\prime&=&-e^{-i\pi/4}W+e^{i\pi/4}\1\\
B^{\prime}&=&i(W+\1).
\end{eqnarray*}
Therefore the S-matrix $S_{\cW}(E)$ for a selfadjoint extension given
in terms of $\cW$, and which has a matrix representation $W$ in the above
bases, takes the form
\begin{eqnarray}\label{n3}
S_{\cW}(E)&=
&-\left(-(e^{-i\pi/4}+\sqrt{E})W+(e^{i\pi/4}-\sqrt{E})\1\right)^{-1}\nonumber\\
&&\cdot\left((-e^{-i\pi/4}+\sqrt{E})W+(e^{i\pi/4}+\sqrt{E})\1\right).
\end{eqnarray}
As a consistency check note that $S_{\cW}(E)$ is unitary for all $E>0$ since
$W$ is unitary. Also $W=-\1$ corresponds to Dirichlet boundary conditions with
$S_{A=\1,B=0}(E)=-\1$ for all $E>0$. This last relation may of course be
inverted to give the matrix $W$ in terms of the S-matrix at any
energy. Indeed
\begin{eqnarray}\label{n4}
W&=&\left((e^{i\pi/4}-\sqrt{E})S(E)+(e^{i\pi/4}+\sqrt{E})\1\right)\nonumber\\
 &&\cdot\left((e^{-i\pi/4}+\sqrt{E})S(E)+(e^{-i\pi/4}-\sqrt{E})\1\right)^{-1}.
\end{eqnarray}
In particular the right hand side is independent of the energy $E$ so this may
serve as another test that the S-matrix indeed results from boundary conditions
in the way described above.

\textbf{Acknowledgements:} The authors would like to thank David
Tomanek for informative discussions concerning nanotubes and an
anonymous referee for helpful critical remarks.

\newpage

\markright{References}

\end{document}